
%
\tolerance=5000
%
\font\xiirm=cmr12       \font\ninerm=cmr9
\font\xiii=cmmi12       \font\ninei=cmmi9
 \skewchar\xiii='177
\font\xiisy=cmsy10 scaled \magstep1
 \skewchar\xiisy='60    \font\ninesy=cmsy9
\font\tenex=cmex10
\font\xiiti=cmti12
\font\xiibf=cmbx12

                        \font\ninebf=cmbx9

\def\xiipoint{\def\rm{\fam0\xiirm}%
  \textfont0=\xiirm    \scriptfont0=\ninerm   \scriptscriptfont0=\sevenrm
  \textfont1=\xiii     \scriptfont1=\ninei    \scriptscriptfont1=\seveni
  \textfont2=\xiisy    \scriptfont2=\ninesy   \scriptscriptfont2=\sevensy
  \textfont3=\tenex    \scriptfont3=\tenex    \scriptscriptfont3=\tenex
  \textfont\itfam=\xiiti                      \def\it{\fam\itfam\xiiti}%
  \textfont\bffam=\xiibf                      \scriptfont\bffam=\ninebf
  \scriptscriptfont\bffam=\sevenbf            \def\bf{\fam\bffam\xiibf}%
  \normalbaselineskip=16.8pt
  \setbox\strutbox=\hbox{\vrule height11.9pt depth4.9pt width0pt}%
  \normalbaselines\rm}
\abovedisplayskip=14pt plus 3.6pt minus 10.8pt
\abovedisplayshortskip=0pt plus 3.6pt
\belowdisplayskip=\abovedisplayskip
\belowdisplayshortskip=8.4pt plus 3.6pt minus 4.8pt
\footline={\hss\xiirm\folio\hss}
\xiipoint
%
\def\underbuildrel#1\over#2{\mathrel{\mathop{\kern0pt #1}\limits_{#2}}}
\def\originalgsim{\underbuildrel>\over\sim}
\def\originallsim{\underbuildrel<\over\sim}
\def\gsim{\mathrel{\raise 2pt\hbox{$\textstyle \originalgsim$}}}
\def\lsim{\mathrel{\raise 2pt\hbox{$\textstyle \originallsim$}}}
\def\scriptgsim{\mathrel{\raise 1pt\hbox{$\scriptscriptstyle \originalgsim$}}}
\def\scriptlsim{\mathrel{\raise 1pt\hbox{$\scriptscriptstyle \originallsim$}}}
\font\smallcap=cmr6
\font\addressfont=cmssi12
\def\sc#1{\hbox{\smallcap #1}}

\def\sss{\scriptscriptstyle}
\def\half{\hbox{$1\over2$}}

\def\num#1.{\item{[#1]}}  \def\bk{\item{}}
\null
\rightline{October, 1993}
\rightline{DPUR 66}
\vskip1.5 true cm
\centerline
{\bf Analytical Solution to the Fokker-Planck Equation with a Bottomless
Action}
\vskip1.5 true cm
\centerline{Hiromichi Nakazato}
\vskip1 true cm
\centerline{\it Department of Physics, University of the Ryukyus}
\centerline{\it Okinawa 903-01, Japan}
\centerline{\vtop{\hbox{\addressfont email: b985039@sci.u-ryukyu.ac.jp}
      \hbox{\hphantom{\addressfont email:}
                        \&\addressfont\ d52787@jpnkudpc.bitnet}}}
\vskip1cm
\centerline{PACS numbers: 11.90.+t, 02.50.Ey}
\vfill
\centerline{\bf Abstract}
\medskip
     A new Langevin equation with a field-dependent kernel is proposed to
deal with bottomless systems within the framework of the stochastic
quantization of Parisi and Wu.
The corresponding Fokker-Planck equation is shown to be a diffusion-type
equation and is solved analytically.
An interesting connection between the solution with the ordinary Feynman
measure, which in this case is not normalizable, is clarified.
\vfill\eject
     One of the interesting and appealing applications of the stochastic
quantization (SQ) of Parisi and Wu [1]
is found in its application to systems characterized by actions unbounded
from below (bottomless systems).
An attempt to stabilize and eventually to quantize bottomless systems is of
great interest in several important fields of physics, such as Euclidean
Einstein gravity.
To deal with bottomless systems on the basis of SQ, Greensite and Halpern [2]
made use of the Fokker-Planck equation w.r.t.\ a fictitious time $t$
$$\eqalignno{
  {\partial\over\partial t}P[\phi;t]&=H[\phi]P[\phi;t],
  &(1\hbox{a})\cr
  H[\phi]&=\int d^D\!x\,
           {\delta\over\delta\phi(x)}\left({\delta\over\delta\phi(x)}
           +{\delta S[\phi]\over\delta\phi(x)}\right),
 &(1\hbox{b})\cr}
$$
with an action $S$ unbounded from below.
In this case, the distribution functional $e^{-S}$ (Feynman measure) cannot
be an equilibrium solution nor even a stationary solution, for it is not
normalizable.
Because the negative semi-definiteness of the Fokker-Planck Hamiltonian $H$
can be proved irrespectively of the boundedness of the action, they proposed
the interesting idea that a possible way to stabilize and to quantize such
bottomless systems in Euclidean space-time is such that the probability
distribution in the Euclidean path integral formulation is given by the
lowest (normalizable) eigenstate of $H$.
This corresponds to using the lowest eigenstate of $H$ of the form
$e^{-S_{\rm eff}}$, instead of $e^{-S}$, as an effective distribution
functional in evaluating expectation values.
It is also shown that such an effective distribution can indeed be extracted
by fixing both initial and final configurations [3].
This procedure avoids runaway solutions which are due to the unboundedness
of the action $S$ of the Langevin equation.
Notice that their proposal for such bottomless systems corresponds to a
departure from the traditional approach that the weight functional in the
path integral formulation be given by $e^{-S}$, with $S$ a classical action.
Remember also that their Langevin equation does not have an equilibrium
distribution because the naive candidate $e^{-S}$ is not normalizable and
hence does not belong to the spectrum of the Fokker-Planck Hamiltonian
$H$: Every eigenstate belongs to its negative-definite eigenvalue.

    Recently Tanaka et al.\ proposed another interesting way of
stochastically quantizing bottomless systems [4].
Unlike the above attempts [2,3], they have pursued the possibility of
reproducing the desired probability distribution $e^{-S}$ in the equilibrium
limit of the fictitious stochastic process even for bottomless systems.
Their starting point is the following Langevin equation
$$
  {\partial\over\partial t}\phi(x,t)
  =-K[\phi]{\delta S[\phi]\over\delta\phi(x,t)}
   +{\delta K[\phi]\over\delta\phi(x,t)}+K^{1\over2}[\phi]\eta(x,t)
  \eqno{(2)}
$$
with a field-dependent positive kernel $K[\phi]$.
Here $\eta$ is a Gaussian white noise with the statistical properties
$$
  \langle\eta(x,t)\rangle=0,\qquad
  \langle\eta(x,t)\eta(x',t')\rangle=2\delta^D\!(x-x')\delta(t-t')
  \eqno{(3)}
$$
and we adopt, here and in what follows, the Ito-interpretation [5]
for stochastic equations.
It is straightforward to derive the corresponding Fokker-Planck equation
from the above Langevin equation (2), and the Fokker-Planck Hamiltonian
is then given by
$$
  H[\phi]=\int d^D\!x\,{\delta\over\delta\phi(x)}K[\phi]
                    \left({\delta\over\delta\phi(x)}
                    +{\delta S[\phi]\over\delta\phi(x)}\right).
  \eqno{(4)}
$$
It is clear from this equation that for any positive definite kernel
functional $K[\phi]$ the spectrum of $H$ is still negative semi-definite so
that the system relaxes to a unique equilibrium state whose distribution is
given by $e^{-S}$, {\it if it is normalizable\/} and if the spectrum has a
discrete zero.
If the classical action $S$ is not bounded from below, the distribution
$e^{-S}$ is not normalizable, and can never be either an equilibrium
solution or a stationary solution of the above equation.
Nevertheless, a better choice of the kernel $K$ may result in the situation
where the stochastic variable $\phi$ is effectively confined to a finite
range, so that the distribution $e^{-S}$ can be regarded as a solution of
(4) normalizable within such a range.
Their choice of the kernel $K$, for example, is
$$
  K[\phi]=e^{-S_4[\phi]}
  \eqno{(5)}
$$
when the action $S$ is given by the difference of two positive monomials
$$\eqalignno{
  &S[\phi]=S_2[\phi]-S_4[\phi],
  &(6\hbox{a})\cr
  &S_2[\phi]={1\over2}\int d^D\!x\,\phi(x)[-\partial^2+m^2]\phi(x),\qquad
  S_4[\phi]={\lambda\over4}\int d^D\!x\,\phi^4(x),\qquad(\lambda>0).
  &(6\hbox{b})\cr}
$$
This choice results in the following form of the Langevin equation
$$
  {\partial\over\partial t}\phi(x,t)
  =-e^{-S_4[\phi]}{\delta S_2[\phi]\over\delta\phi(x,t)}
   +e^{-{1\over2}S_4[\phi]}\eta(x,t),
  \eqno{(7)}
$$
which has a damping drift term.
They have argued [4] that this drift term effectively confines $\phi$ to a
finite range so that the desired distribution $e^{-S}$ can be derived in the
equilibrium limit of (7).
Their numerical simulation of the above kerneled Langevin equation for a
simple zero-dimensional model (wrong-sign $x^4$ model) could indeed reproduce
the distribution $e^{-S(x)}=e^{-{1\over2}m^2x^2+{\lambda\over4}x^4}$ at
equilibrium [4].

     Seeking a possible way of reproducing $e^{-S}$ in the equilibrium limit
for bottomless systems is challenging and very appealing.
However, their proof is crucially dependent on the plausible argument that
the effective range of the stochastic variable may be finite, which makes
the apparently nonnormalizable distribution $e^{-S}$ normalizable, but the
soundness of which is only verified numerically.

     In this short paper, I propose another way of effectively reproducing
$e^{-S}$, starting from the above Langevin equation (2) with a slightly
different kernel $K$.
It will be shown that the resultant Fokker-Planck equation can be solved
exactly at any fictitious time $t<\infty$.
It turns out to be a diffusion-type equation, i.e.\ the spectrum of the
Hamiltonian $H$ is continuous and no equilibrium limit exists.
This might be considered an apparent drawback, but it will also be shown
that the asymptotic form of the probability distribution is well
approximated by the desired form $e^{-S}$ {\it normalized in the finite range
of the stochastic variable, corresponding to a finite fictitious time\/}.
These results, though of course not contradictory to it, form an interesting
contrast with the previous analysis [4], where a discrete spectrum of $H$ is
implicitly assumed and the equilibrium distribution is numerically found to
be well approximated by $e^{-S}$.

     Throughout this paper, we exclusively consider zero-dimensional
potential models whose classical action $S(x)$ is unbounded from below,
for notational simplicity.

     Let us start our discussion with the simplest possible model, i.e.\ the
wrong-sign Gaussian model
$$
  S(x)=-\half mx^2,\qquad (m>0),
  \eqno{(8)}
$$
and show that an exact solution to the Fokker-Planck equation (1a) with the
Hamiltonian (4) is obtainable if we choose an appropriate kernel $K(x)$.
First we divide the above action into two positive monomials as in (6a),
$$
  S(x)=S_0(x)-S_{\sc I}(x),
  \eqno{(9)}
$$
with
$$
  S_0(x)=\half mx^2,\qquad S_{\sc I}(x)=mx^2.
  \eqno{(10)}
$$
Then following Ref.~[4], we are naturally led to a kernel
$$
  K(x)=e^{-S_{\sc I}(x)}=e^{-mx^2}.
  \eqno{(11)}
$$
Though the above separation ((9) and (10)) seems quite arbitrary, it will
be shown that this choice is crucial for the ensuing Fokker-Planck equation
to be solvable and is just an example of a general scheme $K=e^{2S}$ which
in this case is equal to (11) because $2S=-S_{\sc I}$ (see (29) and (32)
below).
The explicit forms of the Langevin and the Fokker-Planck equations are
$$
  \dot x=-e^{-S_{\sc I}(x)}S_0'(x)+e^{-{1\over2}S_{\sc I}(x)}\eta
        =-e^{-mx^2}mx+e^{-{1\over2}mx^2}\eta,
  \eqno{(12)}
$$
and
$$
  \dot P(x;t)={\partial\over\partial x}e^{-S_{\sc I}(x)}
              \left({\partial\over\partial x}+S'(x)\right)P(x;t)
             ={\partial\over\partial x}e^{-mx^2}
              \left({\partial\over\partial x}-mx\right)P(x;t),
  \eqno{(13)}
$$
respectively.
As we know that the corresponding Fokker-Planck Hamiltonian has negative
semi-definite eigenvalues, we put
$$
  P(x;t)=e^{-c^2t}e^{-S(x)}\tilde P(x),
  \eqno{(14)}
$$
with a real parameter $c$.
The above Fokker-Planck equation (13) is then reduced to an ordinary
differential equation for $\tilde P(x)$
$$
  -c^2e^{mx^2}\tilde P(x)=\left({d\over dx}-mx\right){d\over dx}\tilde P(x).
  \eqno{(15)}
$$

     Now we change the variable from $x$ to $f(x)$ in the above equation
and obtain
$$
  \left\{\Bigl(f'(x)\Bigr)^2{d^2\over df^2}
         +\Bigl(f''(x)-mxf'(x)\Bigr){d\over df}+c^2e^{mx^2}\right\}\tilde P(x)
  =0.
  \eqno{(16)}
$$
For the equation to be solvable, we require
$$
  f'(x)\propto e^{mx^2/2}\quad\hbox{and}\quad f''(x)-mxf'(x)=0.
  \eqno{(17)}
$$
It is clear that this condition is identically satisfied for $f'(x)
\propto e^{mx^2/2}$, or
$$
  f(x)=\alpha\int_0^x dy\,e^{my^2/2}+\beta
  \eqno{(18)}
$$
with two real parameters $\alpha$ and $\beta$.
Notice that the transformation between $x$ and the above $f$ is one to one
and the domain of $f$ is again $[-\infty,\infty]$.
This choice of $f(x)$ drastically simplifies the equation (16) and makes
it solvable.
In fact, $\tilde P$ now satisfies
$$
  \left\{\alpha^2{d^2\over df^2}+c^2\right\}\tilde P(x)=0,
  \eqno{(19)}
$$
which is easily solved to give
$$
  \tilde P(x)=\hbox{Re}\Bigl(e^{ic\int_0^x dy\,e^{my^2/2}}\Bigr)\quad
              \hbox{or}\quad
              \hbox{Im}\Bigl(e^{ic\int_0^x dy\,e^{my^2/2}}\Bigr).
  \eqno{(20)}
$$
At this stage it is important to notice that the normalization condition,
which is the only condition that the probability distribution $P$ is subject
to,
$$
  \int_{-\infty}^\infty dx\,P(x;t)
  =\int_{-\infty}^\infty dx\, e^{-c^2t}e^{-S(x)}\tilde P(x)=1
  \eqno{(21)}
$$
seems to be satisfied for any real $c\not=0$ because at large $\vert x\vert$,
which corresponds to large $\vert f\vert$, rapid oscillations of $\tilde P$
(see (20)) may make the integral convergent despite the ever-increasing
factor $e^{-S(x)}$.
No condition has been imposed on the value of $c$, which implies that the
system has a continuous spectrum.
This will be verified shortly.

     Thus we have arrived at the conclusion that the solution of the
Fokker-Planck equation (13) is in general of the form
$$
  P(x;t)=\int_{-\infty}^\infty dc\,e^{-c^2t}e^{-S(x)}
         \hbox{Re}\left[A(c)e^{icf(x)/\alpha}\right]
        =\int_{-\infty}^\infty dc\,e^{-c^2t}e^{-S(x)}
         \hbox{Re}\Biggl[A(c)e^{ic\int_0^x dy\,e^{my^2/2}}\Biggr],
  \eqno{(22)}
$$
where the function $A(c)$ in the second equality has been re-defined to
absorb the constant term in $f(x)$.
Specification of $A(c)$ certainly corresponds to a specific choice of the
initial distribution $P(x;0)$.
It is not difficult to see that the simplest choice of $A(c)$ does
correspond to the most fundamental initial condition.
In fact, if we choose a constant $A=1/2\pi$, we have, at the initial
time $t=0$,
$$
  P(x;0)=\int_{-\infty}^\infty dc\,e^{-S(x)}
          \hbox{Re}\Biggl[{1\over2\pi}
           e^{ic\int_0^x dy\,e^{my^2/2}}\Biggr]
        =e^{mx^2/2}\delta\Bigl[\int_0^x\!dy\,e^{my^2/2}\Bigr]
        =\delta(x).
  \eqno{(23)}
$$
In the last equality we have made use of the one-to-one correspondence
between $x$ and $\int_0^x\!dy\,e^{my^2/2}$ and the monotonically increasing
character of the latter w.r.t.\ $x$.

     In this way, we finally obtain the solution of the Fokker-Planck
equation (13)
$$
  P(x;t)=\int_{-\infty}^\infty dc\,e^{-c^2t}e^{-S(x)}
          \hbox{Re}\Biggl[{1\over2\pi}
           e^{ic\int_0^x dy\,e^{my^2/2}}\Biggr]
        ={1\over\sqrt{4\pi t}}\,e^{-S(x)}
           e^{-{1\over4t}\left(\int_0^x dy\,e^{my^2/2}\right)^2},
  \eqno{(24)}
$$
subject to the initial condition $P(x;0)=\delta(x)$.
It is easily confirmed that this solution is normalizable, i.e.\ that
$\int_{-\infty}^\infty dx\,P(x;t)=1$.
As a trivial generalization, the solution subject to the initial condition
$P(x;0)=P_0(x)$ appears to be
$$
  P(x;t)=\int_{-\infty}^\infty dx_0\,{1\over\sqrt{4\pi t}}\,e^{-S(x)}
          e^{-{1\over4t}\left(\int_{x_0}^x dy\,e^{my^2/2}\right)^2}P_0(x_0).
  \eqno{(25)}
$$
It is instructive to observe that this analytical solution obtained for
the simplest case can be expressed in terms of $S$ only
$$
  P(x;t)=\int_{-\infty}^\infty dx_0\,{1\over\sqrt{4\pi t}}\,e^{-S(x)}
          e^{-{1\over4t}\left(\int_{x_0}^x dy\,e^{-S(y)}\right)^2}P_0(x_0).
  \eqno{(26)}
$$
This suggests that there may in general exist an analytical solution.
This is indeed the case, as will be shown explicitly later.

     It is remarkable that we can obtain analytical solutions to the
Fokker-Planck equation (13), corresponding to the nonlinear Langevin
equation (12).
It should be stressed again that the solution decays at large $t$ and
therefore no equilibrium limit exists and its behaviour is thus quite
different from that expected in the ordinary case: It is rather similar
to that of diffusion processes, which again implies that the spectrum is
continuous.
If, for example, we choose a trivial kernel $K=1$ in (13), no nonlinearity
appears and the standard technique is applicable even though the potential
has the wrong sign [6].
In this case, the solution $P_{\sss K=1}$ subject to $P_{\sss K=1}(x;0)=
\delta(x-x_0)$ is
$$
  P_{\sss K=1}(x;t)=\sqrt{m\over2\pi(1-e^{-2mt})}\,
                     \exp\left[-{m(xe^{-mt}-x_0)^2\over2(1-e^{-2mt})}\right]
                     e^{-mt},
  \eqno{(27)}
$$
whose asymptotic behaviour at large $t$ shows the discreteness of the spectrum
$$
  P_{\sss K=1}(x;t)\buildrel t\to\infty\over\longrightarrow
   \sqrt{m\over2\pi}\,e^{-mx_0^2/2}e^{-mt}.
  \eqno{(28)}
$$

     Now let us turn our attention to more general cases.
No specific form of the classical action $S$ is assumed here, except for its
unboundedness from below.
The problem is to find an appropriate kernel $K$ which makes the
Fokker-Planck equation solvable, as in the previous example.
If we write a positive kernel $K$ as
$$
  K(x)=e^{-W(x)}
  \eqno{(29)}
$$
using a real function $W(x)$, the function $\tilde P$ already introduced
in (14) satisfies
$$
  \left\{\Bigl(f'(x)\Bigr)^2{d^2\over df^2}
         +\Bigl(f''(x)-[W'(x)+S'(x)]f'(x)\Bigr){d\over df}
         +c^2e^{W(x)}\right\}\tilde P(x)
  =0.
  \eqno{(30)}
$$
In this case, solvability requires that
$$
  f'(x)\propto e^{W(x)/2}\quad\hbox{and}\quad f''(x)-[W'(x)+S'(x)]f'(x)=0.
  \eqno{(31)}
$$
These conditions are clearly satisfied by the following $f$ and $W$
$$\eqalignno{
  W(x)&=-2S(x),
  &(32)\cr
  f(x)&=\alpha'\int_0^x dy\,e^{-S(y)}+\beta'.
  &(33)\cr}
$$
Having found the appropriate kernel and the transformation function, we
are therefore able to solve the Fokker-Planck equation, which in this case
is written
$$
  \dot P(x;t)={\partial\over\partial x}e^{2S(x)}
               \left({\partial\over\partial x}+S'(x)\right)P(x;t),
  \eqno{(34)}
$$
just as in the previous example.
The solution $P(x;t)$, satisfying the initial condition $P(x;0)=P_0(x)$,
is straightforwardly found to be
$$
  P(x;t)=\int_{-\infty}^\infty dx_0\,{1\over\sqrt{4\pi t}}\,e^{-S(x)}
          e^{-{1\over4t}\left(\int_{x_0}^x dy\,e^{-S(y)}\right)^2}P_0(x_0),
  \eqno{(35)}
$$
as expected from (26).

     Finally, let us discuss possible implications of these results in
numerical simulations.
One usually keeps track of the stochastic variable, whose time development
is governed by the Langevin equation
$$
  \dot x=e^{2S(x)}S'(x)+e^{S(x)}\eta
  \eqno{(36)}
$$
when the kernel is given by $K=e^{2S}$.
Let the total action $S$ be given by the difference between two terms
$$
  S(x)=S_0(x)-S_{\sc I}(x),
  \eqno{(37)}
$$
where $S_{\sc I}$ is a positive monomial of the highest order, assumed even,
and $S_0(x)$ the remaining terms.
The above Langevin equation is approximated for large $x$ by
$$
  \dot x\sim-e^{-2S_{\sc I}(x)}S_{\sc I}'(x)+e^{-S_{\sc I}(x)}\eta,
  \eqno{(38)}
$$ and for small $x$ by
$$
  \dot x\sim e^{2S_0(x)}S_0'(x)+e^{S_0(x)}\eta.
  \eqno{(39)}
$$
Thus the kernel $K=e^{2S}$ supplies a restoring force for large $x$, but a
diverging force for small $x$.
To understand the behaviour of the stochastic variable $x(t)$ starting from
a finite value $x(0)=x_0$, and the approximate form of the probability
distribution $P$, consider only the finite range $\vert x(t)-x_0\vert\lsim R$.
This range may be viewed as an actual domain of the variable in numerical
simulations;
(no infinite range can be realized in practical simulations.)
Next we define $t_{\sc R}$ as a time scale satisfying
$$
  {1\over4t_{\sc R}}\left(\int_{x_0-R}^{x_0+R}dy\,e^{-S(y)}\right)^2=1.
  \eqno{(40)}
$$
Then at $t=t_{\sc R}$, the second exponential factor of the exact solution
(35) is expressed as
$$
  e^{-{1\over4t_{\sc R}}\left(\int_{x_0}^x dy\,e^{-S(y)}\right)^2}
  =e^{-\left(\int_{x_0}^x dy\,e^{-S(y)}\right)^2\Bigl/
           \left(\int_{x_0-R}^{x_0+R}dy\,e^{-S(y)}\right)^2}
  \eqno{(41)}
$$
and is considered to be of the order of unity for any $x$ in the range
$\vert x-x_0\vert\lsim R$.
Under this estimation, the probability distribution at time $t_{\sc R}$ is
well approximated by a naive distribution $e^{-S}$ normalized in this range
$$
  P(x;t_{\sc R})\sim{e^{-S(x)}\over Z_R},\qquad
  Z_R=\int_{x_0-R}^{x_0+R}dy\,e^{-S(y)}.
  \eqno{(42)}
$$
This implies the possibility in the actual simulation of the Langevin
equation (36) of reproducing the desired probability distribution $e^{-S}$
at a certain finite time $t_{\sc R}$, normalized in a finite domain
determined by the details of the simulation.
Further investigation is necessary to draw more definite conclusions.
\bigskip

     The author acknowledges useful and informative discussions with
Profs.\ I.~Ohba, Y.~Yamanaka and K.~Okano and Drs. S.~Tanaka, M.~Mizutani
and M.~Kanenaga.
\vfill
\eject
\noindent
{\bf References}
\medskip
\num1.  G. Parisi and Y.-S. Wu, Sci.\ Sin.\ {\bf 24} (1981) 483.
\bk     For review articles, see
\bk     P. H. Damgaard and H. H\"uffel, Phys.\ Rep.\ {\bf 152} (1987) 227;
\bk     M. Namiki, {\it Stochastic Quantization\/}, (Springer-Verlag,
        Heidelberg, 1992).
\num2.  J. Greensite and M. B. Halpern, Nucl.\ Phys.\ {\bf B242} (1984) 167.
\num3.  J. Greensite, Nucl.\ Phys.\ {\bf B439} (1993) 439.
\num4.  S. Tanaka, M. Namiki, I. Ohba, M. Mizutani, N. Komoike and M.
        Kanenaga, Phys.\ Lett.\ {\bf B288} (1992) 129.
\bk     S. Tanaka, I. Ohba, M. Namiki, M. Mizutani, N. Komoike and M.
        Kanenaga, Prog.\ Theor.\ Phys.\ {\bf 89} (1993) 187.
\bk     M. Kanenaga, M. Mizutani, M. Namiki, I. Ohba and S. Tanaka, Waseda
        Univ.\ preprint WU-HEP-92-8.
\num5.  See for example,
\bk     C. G. Gardiner, {\it Handbook of Stochastic Methods for Physics,
        Chemistry and the Natural Sciences\/}, (Springer-Verlag, New York,
        1985).
\num6.  For inverted-potential problems in stochastic processes, see
\bk     H. Risken, {\it The Fokker-Planck Equation\/}, (Springer-Verlag,
        New York, 1989).
\bye